\documentclass[runningheads]{llncs}
\usepackage{dialogue}
\usepackage{tabularx}
\usepackage{amssymb}
\usepackage{amsmath} 
\usepackage{algorithm}
\usepackage{algpseudocode}
\usepackage{float}
\usepackage{mdframed}
\usepackage{xcolor}
\usepackage{threeparttable}
\usepackage[T1]{fontenc}
\usepackage{graphicx}
\usepackage{enumitem}
\usepackage{subcaption}
\usepackage{orcidlink}
\begin{document}
\title{Framing the 5\% Problem: Teachers' Perspectives on Persistence in Educational Technology}
\titlerunning{Teachers' Perspectives on the 5\% Problem}
\author{Conrad Borchers\orcidlink{0000-0003-3437-8979}}
\institute{
Vanderbilt University, Nashville, TN, USA\\
\email{c.borchers@vanderbilt.edu}
}
\authorrunning{C. Borchers}
\maketitle 
\begin{abstract}
Adaptive K–12 mathematics platforms often show low sustained student use, a pattern termed the 5\% problem. Although prior work has developed analytics and interventions to identify disengagement, less is known about how teachers make sense of persistence challenges in individualized learning classrooms. This study reports on a 90-minute participatory design workshop with 12 U.S. middle school mathematics teachers who use i-Ready Math weekly. Teachers generated needs through open-ended prompts, card sorting, and collective voting. Thematic analysis identified four recurring dimensions of low persistence spanning motivation and buy-in, cognitive roadblocks, resilience under challenge, and contextual barriers. Teachers emphasized the need to identify where students become stuck and recognize silent disengagement, prioritizing support for diagnosis and timely instructional response over aggregate usage metrics. These findings reframe the 5\% problem as a situated instructional challenge and suggest a need for teacher-facing systems that support interpretation of student persistence through both cognitive and contextual evidence.
\end{abstract}
\keywords{participatory design \and K-12 \and mathematics education \and persistence}

\section{Introduction \& Related Work}

Intelligent tutoring systems have demonstrated strong evidence of supporting student learning~\cite{kulik2016effectiveness}. Yet at scale, learning platforms consistently exhibit what has been termed the \emph{5\% problem}: only a small fraction of students engage at levels considered sufficient for meaningful learning gains, often benchmarked at 30 minutes per week~\cite{holt20245}. This pattern has been documented across platforms including Khan Academy and IXL~\cite{holt20245}. Recent evidence from Khan Academy further illustrates the challenge, showing that classrooms averaged about 11 minutes of weekly practice (roughly 6.6 hours per year), despite projected learning gains increasing roughly linearly through commonly recommended usage targets~\cite{eames2026computer}.

The importance of student persistence has similarly been a long-standing topic of interest to the field of technology-enhanced learning. Complementary lines of relevant work illustrate how engagement and persistence are often pursued through student-facing mechanisms (e.g., goal-setting with rewards \cite{borchers2025engagement}), technology-mediated nudging (e.g., through analytics-based early alerts that send automated e-mail and SMS prompts when LMS use flags disengagement in higher education \cite{kay2023power}), and gamification-oriented designs as discussed in literature reviews \cite{christopoulos2023gamification}. These contributions are valuable, yet they do not, by themselves, specify what teachers need in order to interpret heterogeneous forms of disengagement when platforms are embedded in everyday classroom routines. Teachers, who orchestrate technology use during instruction, frequently lack visibility into \textit{why} students disengage, \textit{where} they become stuck, and \textit{which} barriers are amenable to teacher support \cite{holstein2019co,yang2024leveraging}. At the same time, prior qualitative and mixed-methods research shows that outcomes with learning software are shaped as much by implementation conditions and teacher capacity as by platform design, raising equity concerns when these contextual factors are ignored \cite{heinrich2020equity}.

Recent EC-TEL research has increasingly examined how teachers interpret and act on data in technology-mediated classrooms. Yang et al. investigate how teachers balance the value of multimodal classroom evidence for reflection against concerns about privacy and manageability \cite{yang2024leveraging}. Kasepalu et al. emphasize the role of situation-specific professional competencies in enabling teachers to act on learning analytics \cite{kasepalu2024situation}, while Nezhad et al. identify middle-school mathematics teachers' information needs when coordinating student work with digital practice \cite{nezhad2024math}. Nazaretsky argues that participatory design should be linked to educator professional development \cite{nazaretsky2025educator}, and Falhs et al. show that teacher expertise shapes preferences for AI reflection supports \cite{falhs2025reflection}. Building on these efforts, this study examines the 5\% problem, focusing on how teachers conceptualize disengagement and what forms of support they prioritize.

To illuminate how teachers conceptualize student persistence, this study employed a participatory design workshop with middle school mathematics teachers who use i-Ready Math at least weekly. i-Ready Math is an adaptive learning platform that combines diagnostic assessment with personalized mathematics instruction and provides teachers with analytics reporting student progress, content mastery, and time-on-task. Through structured activities focused on imagining instructional capabilities and collectively prioritizing persistence challenges, the workshop elicited teachers' perspectives on student effort, disengagement, and persistence. The study addresses the following research questions:

\begin{itemize}[leftmargin=*]
\item \textbf{RQ1:} How do middle school mathematics teachers conceptualize the causes of student disengagement and lack of persistence when using i-Ready Math?
\item \textbf{RQ2:} Which dimensions of the 5\% problem do teachers collectively identify as most critical for supporting student persistence?
\end{itemize}

\section{Methods}

The study employed participatory design needs-finding to surface teacher perspectives on persistence in computer-assisted learning \cite{holstein2019co}. Twelve middle school mathematics teachers from two schools in one public school district in the Pacific United States participated voluntarily. All used i-Ready Math at least weekly. The district serves a socioeconomically mixed urban community. Teachers received \$125 for a 90-minute session approved by an institutional review board.

The workshop opened with a shared scenario: students access i-Ready Math one period per week yet average only a few minutes of practice. A ``superpower'' elicitation activity adapted from prior teacher co-design work \cite{holstein2019co} prompted ``I wish I knew\ldots'' and ``I wish I could\ldots'' responses. Teachers first brainstormed independently on index cards for 20 minutes, then clustered cards in groups of 2-4 through open card sorting \cite{wood2008card} while audio recording discussion. Each group named its top challenge, posted a summary, and the full cohort dot-voted on five stickers per person. A brief whole-group reflection closed the session. Data comprised individual cards, grouped hierarchies, voting tallies, and recordings.

Two researchers conducted reflexive thematic analysis of cards, group artifacts, voting outcomes, and discussion transcripts \cite{terry2017thematic}. Each researcher independently open-coded cards and transcript segments for references to causes of disengagement, barriers to persistence, and desired teacher capabilities. Codes were then compared and consolidated through discussion and the categories generated during group card sorting. Candidate themes were developed by grouping related codes and merging overlapping categories identified across researcher notes. For example, references to attendance, device access, and other external factors were grouped as contextual barriers. Theme labels were chosen to reflect teacher language where possible, including terms such as buy-in and roadblock. Voting results were used to contextualize the relative salience of themes. Final themes were retained when supported by multiple forms of evidence.

\section{Results}

\begin{figure}[htpb]
  \centering
  \begin{subfigure}[t]{0.45\linewidth}
    \centering
    \includegraphics[width=\linewidth]{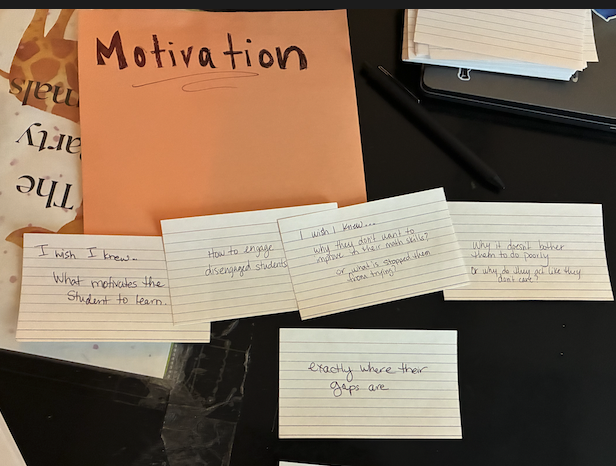}
    \caption{Card sorting hierarchy activity results (example from a single group).}
    \label{fig:hierarchy}
  \end{subfigure}\hfill
  \begin{subfigure}[t]{0.45\linewidth}
    \centering
    \includegraphics[width=\linewidth]{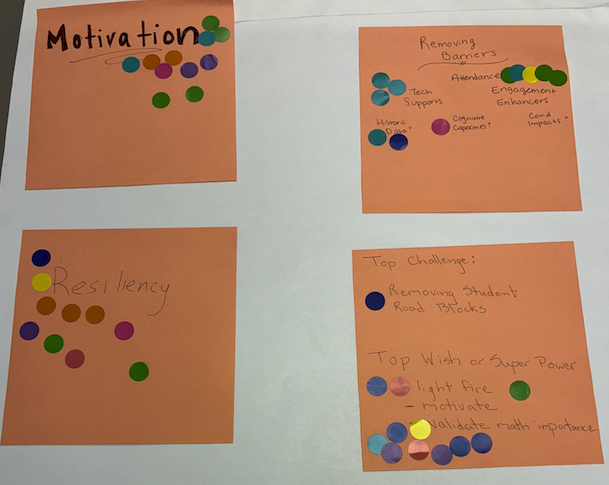}
    \caption{Voting activity results on teacher group summaries.}
    \label{fig:voting}
  \end{subfigure}
  \caption{Workshop artifacts: (a) hierarchy mapping and (b) dot-voting outcomes.}
  \label{fig:artifacts_side_by_side}
\end{figure}

Teachers framed low persistence as an interaction among motivational, cognitive, and contextual factors (Figure~\ref{fig:artifacts_side_by_side}). Four themes recurred across workshop activities. \textbf{Motivation and buy-in.} Teachers described students who ``do not care,'' comply without effort, or disengage without visible frustration. They linked motivation to middle-school developmental constraints, peer norms, and prior negative experiences with math or i-Ready. \textbf{Stuck points and cognitive roadblocks.} Teachers reported difficulty locating specific conceptual gaps and distinguishing confusion from avoidance. Students often stopped progressing silently after difficulty, which made disengagement hard to detect during class. \textbf{Resilience and persistence under challenge.} Teachers described students who froze, abandoned problems after initial failure, or lacked independent problem-solving strategies. They treated resilience as a teachable skill that current routines and dashboards did little to support. \textbf{Contextual and structural barriers.} Attendance, unreliable devices, prior academic gaps, and COVID-related disruption appeared as compounding external constraints.

Dot voting showed no single dominant category. Instead, three interrelated priorities aligned with the themes above: motivation and buy-in, removing student roadblocks, and building resilience under challenge. Teachers paired each priority with a corresponding ``superpower,'' such as knowing where students stall or reframing productive struggle. They emphasized interpretive needs to understand students' context, detecting silent disengagement, distinguishing confusion from avoidance and deciding when to intervene.

\section{Discussion}

This study examined the 5\% problem from a teacher-centered perspective. Addressing RQ1, teachers consistently framed disengagement as a situated instructional challenge. Consistent with prior work on teacher sensemaking around educational technology \cite{holstein2019co}, participants emphasized uncertainty about why students disengage and which barriers are amenable to instructional support. For instance, teachers distinguished between overt disengagement and students who appeared compliant while investing minimal cognitive effort. This distinction aligns with research on teacher noticing in tutoring environments \cite{karumbaiah2023spatiotemporal}. Together, these findings suggest that the 5\% problem extends beyond time-on-task and reflects challenges in interpreting meaningful student effort \cite{holt20245}. For RQ2, motivation, roadblocks, and resilience functioned as mutually reinforcing dimensions: unidentified roadblocks fed disengagement, which weakened buy-in, and limited resilience amplified both. Analytics for persistence should therefore support how teachers reason about disengagement as it unfolds \cite{nezhad2024math,yang2024leveraging}.

A key implication concerns the limits of existing platform data for understanding the contextual factors that teachers identified as shaping persistence. By extension, these findings suggest limits to prevailing conceptions of the 5\% problem that rely primarily on time-based platform usage patterns. Consistent with prior work showing that implementation conditions influence educational technology use \cite{heinrich2020equity}, participants expressed uncertainty about whether disengagement reflected conceptual difficulty, motivation, attendance, prior experiences with mathematics, or circumstances beyond the platform itself. Accordingly, teacher-facing systems may need to support interpretation by situating engagement patterns within broader instructional and contextual information. This points toward analytics that support teacher sensemaking and instructional decision-making around persistence, complementing prior work in learning analytics that has primarily emphasized the detection of disengagement \cite{holstein2019co,baker2011gaming}.

As limitations worth noting, this study is an early needs-finding workshop with 12 volunteer teachers, one 90-minute session, one district, and one platform. Perspectives may reflect local implementation norms and partnership recruitment. Group discussion may have amplified vocal participants, and card-sort labels partially shaped later coding. Triangulation with classroom observations, student interviews, or longitudinal prototype trials would test transferability.

\section{Conclusion}

This study examined the 5\% problem from a teacher perspective through a participatory design workshop with middle school math teachers who use i-Ready Math. Teachers described low persistence in terms of motivation and buy-in, cognitive roadblocks, resilience under challenge, and contextual barriers. They prioritized support for identifying where students become stuck, recognizing silent disengagement, and informing intervention. These findings suggest that the 5\% problem cannot be understood through platform usage patterns alone. For the field, they highlight opportunities for teacher-facing systems that support interpretation of persistence within classroom contexts. Future work will explore these opportunities through participatory prototyping and classroom evaluation.

\section*{Acknowledgments}

This work was supported, in whole or in part, by the Bill \& Melinda Gates Foundation INV-068909. The conclusions and opinions expressed in this work are those of the author alone and shall not be attributed to the Foundation. Under the Foundation's grant conditions, a Creative Commons Attribution 4.0 Generic License has already been assigned to the Author Accepted Manuscript.

\bibliographystyle{splncs04}
\bibliography{main}

\end{document}